\documentclass[prd,superscriptaddress,twocolumn,nofootinbib]{revtex4-2}
\usepackage{color}
\usepackage{graphicx}
\usepackage{amsfonts}
\usepackage{amssymb}
\usepackage{amsbsy}
\usepackage{amsmath}
\usepackage{mathrsfs}
\usepackage{latexsym}
\usepackage{natbib}
\usepackage{bm}
\usepackage{subfigure}
\usepackage{color}
\usepackage{dirtytalk}
\usepackage{hyperref}
\usepackage{float}

\definecolor{LinkColor}{rgb}{0.75, 0, 0}
\definecolor{CiteColor}{rgb}{0, 0.5, 0.5}
\definecolor{UrlColor}{rgb}{0, 0, 0.75}
\hypersetup{linkcolor=LinkColor}
\hypersetup{citecolor=CiteColor}
\hypersetup{urlcolor=UrlColor}

\begin{document} 
\title{Hawking radiation as quantum mechanical reflection}

\author{Pritam Nanda}
\email{pritam.nanda@saha.ac.in}
\affiliation{%
 Theory Division, Saha Institute of Nuclear Physics. 1/AF Bidhan Nagar, Kolkata 700064, India and Homi Bhabha National Institute, Training School Complex, Anushaktinagar, Mumbai
 400094, India}

\author{Chiranjeeb Singha}
 \email{chiranjeeb.singha@saha.ac.in}
\affiliation{%
 Theory Division, Saha Institute of Nuclear Physics. 1/AF Bidhan Nagar, Kolkata 700064,}
 
 \author{Pabitra Tripathy}
\email{pabitra.tripathy@saha.ac.in}
\affiliation{%
 Theory Division, Saha Institute of Nuclear Physics. 1/AF Bidhan Nagar, Kolkata 700064, India and Homi Bhabha National Institute, Training School Complex, Anushaktinagar, Mumbai
 400094, India}
 
 \author{Amit Ghosh}
\email{amit.ghosh@saha.ac.in}
 \affiliation{%
 Theory Division, Saha Institute of Nuclear Physics. 1/AF Bidhan Nagar, Kolkata 700064, India and Homi Bhabha National Institute, Training School Complex, Anushaktinagar, Mumbai
400094, India}


\date{\today}

\begin{abstract}

In this article, we explore an alternative derivation of Hawking radiation. Instead of the field-theoretic derivation, we have suggested a simpler calculation based on quantum mechanical reflection from a one-dimensional potential. The reflection coefficient shows an exponential fall in energy which, in comparison with the Boltzmann probability distribution, yields a temperature. The temperature is the same as Hawking temperature for spherically symmetric black holes. The derivation gives an exact local calculation of Hawking temperature that involves a region lying entirely outside the horizon. This is a crucial difference from the tunneling calculation, where it is necessary to involve a region inside the horizon.
 
\end{abstract}

\maketitle


\section{INTRODUCTION}

Black holes are the simplest solution of the Einstein field equation. Classically, it is defined as the region of spacetime from which no information can reach the asymptotic infinity $\mathscr I^+$ \cite{Hawking:1971vc}. However, Hawking showed that although a classical black hole is completely black, quantum-mechanically, it can emit particles at a temperature \cite{Hawking:1974sw}. Hawking's work was based on the scattering of quantum fields in a classical black hole background. It was shown that if we prepare an initial vacuum state at past null infinity $\mathscr I^-$ and allow it to evolve it in the background of a black hole, in particular in a collapsing spacetime, the final state would be a thermal state at future null infinity $\mathscr I^+$. This is interpreted as if the black hole is emitting thermal radiation, causing its mass to decrease gradually, along with the angular momentum and electric charge. The original calculations of Hawking were extended by Wald to full-scale quantum field theory calculations in \cite{Wald:1975kc}. One of the main characteristics of the original calculation is that it makes use of the global properties of spacetime. The whole work is based on the notion of the event horizon. To define an event horizon, the knowledge of the asymptotic boundary of spacetime is needed \cite{Hawking73,wald2010general,Ashtekar:2004cn}. So an observer should wait for an infinite time to observe the Hawking radiation. To bypass this issue, recently, a lot of work has been carried out to derive Hawking radiation using local properties of spacetime. Generally speaking, a black hole is best described by its horizon, not by the asymptotic boundary. So it is reasonable to expect that the Hawking radiation can be derived using near-horizon properties of spacetime alone. There is a plethora of recent work that made investigations in this direction \cite{Parikh:1999mf,Angheben:2005rm,Robinson:2005pd,Chatterjee:2007hc,Shankaranarayanan:2000qv,Angheben:2005rm,Kerner:2006vu,1973JETP...37...28S,1968JMP.....9..163M,PhysRevD.14.332,Sannan:1988eh,PhysRevD.60.024007,PhysRevD.36.1269}. Among all these works, tunneling calculations \cite{Parikh:1999mf,Akhmedov:2006pg} come with an interesting framework for deriving the Hawking radiation. In this framework, the process of Hawking radiation is realized as pair creation near or inside the horizon. As one particle of the pair crosses the horizon by tunneling, the other particle escapes to infinity. The particle falling into the black hole carries negative energy, while the one that escapes to infinity carries positive energy so that total energy remains zero. Following this energy conservation, it was argued in \cite{Parikh:1999mf} that the black hole must reduce in mass through radiation. The main problem in tunneling formalism is that we have to define particle states inside the horizon. As the time-like Killing coordinate inside the horizon become space-like, the energy of the particle is not well-defined.

In this paper, we follow the opposite approach. Instead of tunneling, we map the whole problem to a quantum mechanical above-barrier-reflection problem \cite{book:14848,2010} from a one-dimensional potential well. We figured out that if we write a free massless scalar field equation in a spherically symmetric black hole background or an axisymmetric black hole background, it looks like a Schrödinger equation for a unit mass particle in a potential generated due to the curvature of spacetime. Using the standard quantum mechanics techniques, we calculate the reflection coefficient from this potential Barrier. And we find that the reflection coefficient has an exponential fall-off with the energy of the particle.

In calculating the reflection coefficient, we have adopted two approaches, one is given in section \ref{Schrodinger}, and another is in section \ref{exactsolution}. In section \ref{Schrodinger}, we have mapped the exact one-dimensional Schrödinger equation to an approximate Schrödinger equation by making a near horizon expansion of the radial coordinate. Then by solving the approximate Schrödinger equation, we obtain the reflection coefficient. To give a strong support in favour of the approximation used in the section \ref{Schrodinger}, in section \ref{exactsolution}, we have dealt with the exact Schrödinger equation whose kinetic part is given in Eddington-Finkelstein coordinate $r_{\star}$ however the potential part is given in coordinate $r$. So we approximate potential, which is Poschl-Teller potential with parameters reproducing the original potential for a wide range of $r_{\star}$. Then the reflection coefficient would be obtained by Poschl-Teller potential. The two temperatures are differed by $\sqrt{2}$. A brief conclusion is added at the end in Sec. \ref{conclusion}.

We have used natural unit $c = G =\hbar= 1$ throughout the paper.

\section{Recovering the near-horizon Schrödinger equation}\label{Schrodinger}

We consider a free massless scalar field $\phi(x)$ in a fixed curved background with a metric $g_{\mu\nu}$.
The dynamics of the scalar field is governed by the action \cite{Bire82}
\begin{equation}\label{eqn:1}
S=-\frac{1}{2}\int\sqrt{-g}d^4x(g^{\mu\nu}\nabla_\mu\phi\nabla_\nu\phi)~.
\end{equation}
The variation of the action with respect to the metric $g_{\mu\nu}$ gives the equation of motion which is the Klein-Gordon equation in a curved background,
\begin{equation}\label{eqn:2}
 \frac{1}{\sqrt{-g}}\partial_\mu(\sqrt{-g}g^{\mu\nu}\partial_\nu\phi)=0~.   
\end{equation}
We also consider a spherically symmetric black hole with the metric signature $(-,+,+,+)$,
\begin{equation}\label{eqn:3}
 ds^2=-f(r)dt^2+\frac{dr^2}{f(r)}+r^2d\Omega^2~,
\end{equation}
where $t$ and $r$ are the time and radial coordinates respectively, and $d\Omega^2= d\theta^2+\sin^2\theta d\phi^2$ is the two-sphere metric. From the separation of the equation of motion 
(\ref{eqn:2})
in the background metric
(\ref{eqn:3}), we can extract the radial equation as,
\begin{equation}\label{eqn:4}
-\partial_t^2\phi+\frac{f(r)}{r^2}\partial_r(r^2f(r)\partial_r\phi)=0~.
\end{equation}
By isolating a specific mode of the scalar field, $\phi(t,r)=e^{i\omega t}\frac{\phi(r)}{r}$ and making use of the coordinate transformation  $dr_*=\frac{dr}{f(r)}$, the radial Klein-Gordon equation takes the following form
\begin{equation}\label{eqn:5}
\partial^2_{r_*}\phi+\left(\omega^2-\frac{f^\prime(r)f(r)}{r}\right)\phi=0~.
\end{equation}
It is to be noted that this is a linear, homogeneous, second-order differential equation, which is similar in form to the Schrödinger equation of quantum mechanics (QM) in one-dimension. Here `prime' denotes the derivative with respect to $r$. However, this equation has been obtained from a field theory and not from a quantum mechanical potential, is reflected in the energy term of the equation --- we find the energy term to be energy-square, not simply energy as would be expected in QM. Because we are dealing with a relativistic field equation, the analog Schrödinger equation bears the characteristic features of the field theory. In field theory, the energy of a one-particle state $a_k^\dagger e^{i{\bf k\cdot x}}|0\rangle\sim|\bf x,k\rangle$ is $\omega=|\bf k|$. The integral operator acting on the vacuum is nothing but the Fourier decomposition of the field operator $\phi({\bf x,k})$. So the analogy of (\ref{eqn:5}) with the Schrödinger equation for a one-particle state is well-justified. 

\subsection{Solution of Schrödinger equation near horizon}

In general it is not always possible to find the explicit relation between r and $r_*$, so it is plausible to solve 
Eqn. (\ref{eqn:5})
in r coordinate. We express the Schrödinger equation in r coordinate . Start with
Eqn. (\ref{eqn:4})
and take the separable form of the scalar field as, $\phi(r,t)=e^{i\omega t}\phi(r)$. The resultant equation is as follows,
\begin{equation}\label{eqn:6}
\partial_r^2\phi+\left(\frac{f^\prime(r)}{f(r)}+\frac{2}{r}\right)\partial_r\phi+\frac{\omega^2}{f(r)^2}\phi=0~.    
\end{equation}
If we start with a black hole space-time then $f(r)$ has a zero at the horizon radius $r=r_0$ and we expand it near horizon. Let, $r=r_0+\epsilon$ where $\epsilon$ is much much less than $r_0$. Now we can make a Taylor's series expansion of $f(r)$ with respect to $r_0$ as,
\[\begin{split}
f(r)=f(r_0)+(r-r_0)f^\prime(r_0)+\frac{1}{2}(r-r_0)^2f^{\prime\prime}(r_0)+\\\frac{1}{6}(r-r_0)^3f^{\prime\prime\prime}(r_0)+ higher\: order~.    
\end{split}\] 
By introducing the surface gravity term $\kappa=\frac{1}{2}f^\prime (r_0)$ and after simplification $f(r)$ can be rewritten as
\begin{equation}
f(r)=2\kappa\epsilon+\frac{1}{2}f^{\prime\prime}(r_0)\epsilon^2+\frac{1}{6}f^{\prime\prime\prime}(r_0)\epsilon^3+\mathcal{O}(\epsilon^4)~.
\end{equation}
For near horizon region($\epsilon\rightarrow0$) we keep terms up to $\epsilon^3$ in $f(r)$. Then the 
Eqn. (\ref{eqn:6})
can be written in terms of $\epsilon$ as,
\begin{multline}\label{eqn:7}
\partial_\epsilon^2\phi+\frac{1}{\epsilon}\partial_\epsilon\phi+\omega^2\frac{1}{4\kappa^2}\left(\frac{1}{\epsilon^2}-\frac{f^{\prime\prime}(r_0)}{2\kappa}\frac{1}{\epsilon}\right.\\\left.+3\frac{f^{\prime\prime}(r_0)^2}{16\kappa^2}-\frac{f^{\prime\prime\prime}(r_0)}{6\kappa}\right)\phi=0~.    
\end{multline}
Now we transform $\phi$ as $\phi\rightarrow e^{-\frac{1}{2}\log\epsilon}\psi$ and consider $\frac{\omega}{\kappa}>>1$.
By imposing all the above conditions, 
Eqn. (\ref{eqn:7})
becomes,
\begin{multline}\label{eq9}
\partial_\epsilon^2\psi+\frac{\omega^2}{4\kappa^2}\left(\frac{1}{\epsilon^2}-\frac{f^{\prime\prime}(r_0)}{2\kappa}\frac{1}{\epsilon}+\right.\\\left.\left(3\frac{f^{\prime\prime}(r_0)^2}{16\kappa^2}-\frac{f^{\prime\prime\prime}(r_0)}{6\kappa}\right)\right)\psi=0~.
\end{multline}
To simplify the above equation, we define the coefficients as follows
\begin{equation}
\begin{split}
A &=\frac{1}{4\kappa^2}\left(\frac{3f^{\prime\prime}(r_0)^2}{16\kappa^2}-\frac{f^{\prime\prime\prime}(r_0)}{6\kappa}\right)~,\\B &=-\frac{1}{8\kappa^3}f^{\prime\prime}(r_0)~,\\C&=\frac{1}{4\kappa^2}~.
\end{split}
\end{equation}
In terms of A, B and C Eqn. (\ref{eq9})  can be written as
\begin{equation}\label{eq11}
\partial_\epsilon^2\psi+\omega^2\bigg(A+\frac{B}{\epsilon}+\frac{C}{\epsilon^2}\bigg)\psi=0~.
\end{equation}
One can get the solution of Eqn. (\ref{eq11}) as,
\begin{equation}
\begin{split}
\psi(\epsilon)=&C_1 e^{(-i\omega\sqrt{A}\epsilon)}(2i\omega\sqrt{A}\epsilon)^{-i\sqrt{C}\omega+\frac{1}{2}}\times\\ &M\left(\frac{1}{2}-i\sqrt{C}\omega+\frac{iB\omega}{2\sqrt{A}},1-2i\sqrt{C}\omega,2i\omega\sqrt{A}\epsilon\right)+\\&C_2e^{(-i\omega\sqrt{A}\epsilon)}(2i\omega\sqrt{A}\epsilon)^{-i\sqrt{C}\omega+\frac{1}{2}}\times\\&U\left(\frac{1}{2}-i\sqrt{C}\omega+\frac{iB\omega}{2\sqrt{A}},1-2i\sqrt{C}\omega,2i\omega\sqrt{A}\epsilon\right)~.
\end{split}
\end{equation}
Finally, the solution of Eqn. (\ref{eq11}) in terms of $\phi$ can be expressed as,
\begin{equation}\label{eq13}
\begin{split}
\phi(\epsilon)=&C_1 (2i\omega\sqrt{A}\epsilon)^{-i\sqrt{C}\omega}\times\\ &M\left(\frac{1}{2}-i\sqrt{C}\omega+\frac{iB\omega}{2\sqrt{A}},1-2i\sqrt{C}\omega,2i\omega\sqrt{A}\epsilon\right)\\&+C_2(2i\omega\sqrt{A}\epsilon)^{-i\sqrt{C}\omega}\times\\&U\left(\frac{1}{2}-i\sqrt{C}\omega+\frac{iB\omega}{2\sqrt{A}},1-2i\sqrt{C}\omega,2i\omega\sqrt{A}\epsilon\right)~.
\end{split}
\end{equation}
Using the asymptotic expression of confluent hyper-geometric function Appendix \ref{appendix} for $(\epsilon\rightarrow0)$, we could simplify the above solution as, 
\begin{equation}\label{eq14}
\begin{split}
\phi(\epsilon)=&C_1 (2i\omega\sqrt{A}\epsilon)^{-i\sqrt{C}\omega} \\&+C_2(2i\omega\sqrt{A}\epsilon)^{i\sqrt{C}\omega}\frac{\Gamma(-2i\sqrt{C}\omega)}{\Gamma\left(\frac{1}{2}-i\sqrt{C}\omega+\frac{iB\omega}{2\sqrt{A}}\right)}~.
\end{split}
\end{equation}

\subsection{Calculation of reflection coefficient}
To calculate the reflection coefficient we use the simple technique of quantum mechanics. Near horizon, the $r_*$ behave as $\log(\epsilon)$ so the expression $\epsilon^{\pm i\omega}$ could be expressed as $e^{\pm i\omega r_*}$ . Now the solution of the Eqn. (\ref{eq14}) could be easily expressed as a linear combination of incoming and outgoing wave function as, 
\begin{equation}
    \phi(r_*)=\mathcal{A}e^{-i\omega r_*}+\mathcal{B}e^{+i\omega r_*}~,
\end{equation}
where $\mathcal{A}(\omega)$ and $\mathcal{B}(\omega)$ is defined as follows,
\begin{equation}
\begin{split}
    &\mathcal{A}(\omega)=C_1 (2i\omega\sqrt{A})^{-i\sqrt{C}\omega}~,\\
    &\mathcal{B}(\omega)=C_2(2i\omega\sqrt{A})^{i\sqrt{C}\omega}\frac{\Gamma(-2i\sqrt{C}\omega)}{\Gamma\left(\frac{1}{2}-i\sqrt{C}\omega+\frac{iB\omega}{2\sqrt{A}}\right)}~.
\end{split}
\end{equation}
The reflection coefficient is defined as \cite{book:14848, Gogberashvili:2017xti, Gogberashvili:2016xcx},
\begin{equation}
\mathcal{R}=\frac{\mathcal{B}\mathcal{B^*}}{\mathcal{A}\mathcal{A^*}}~.
\end{equation}
Using the above definition of reflection coefficient, we calculate it for the solution Eqn. (\ref{eq14})  as,
\begin{equation}
R(\omega)=\mathcal{C}(\omega)\frac{\Gamma(-2i\sqrt{C}\omega)\Gamma(2i\sqrt{C}\omega)~e^{-2\pi\sqrt{C}\omega}}{\Gamma\left(\frac{1}{2}-i\sqrt{C}\omega+\frac{i B\omega}{2\sqrt{A}}\right)\Gamma\left(\frac{1}{2}+i\sqrt{C}\omega-\frac{i B\omega}{2\sqrt{A}}\right)}~,
\end{equation}
where $\mathcal{C}(\omega)=\frac{\left| C_1\right|^2}{\left| C_2\right|^2 }$. Now using some well-known identities such as 
\begin{equation}\label{identities}
\begin{split}
&\Gamma(i y)\Gamma(-i y)=\frac{\pi}{y\sinh(\pi y)} \;\;\; \textrm{and}\\
&\Gamma \left(\frac{1}{2}+iy\right)\Gamma \left(\frac{1}{2}-iy\right)=\frac{\pi}{\cosh(\pi y)}~,
\end{split}
\end{equation}
we could simplify the expression of reflection coefficient as, 

\begin{equation}
R(\omega)=\mathcal{C}(\omega)\frac{\cosh{\left(\pi\left(\sqrt{C}\omega-\frac{B\omega}{2\sqrt{A}}\right)\right)}}{2\sqrt{C}\omega\sinh({2\pi\sqrt{C}\omega)}}e^{-2\pi\sqrt{C}\omega}~,
\end{equation}
and also, using the condition that the energy of the particle is much higher than curvature $(\frac{\omega}{\kappa}>>1)$ we could write the reflection coefficient as,
\begin{equation}
R(\omega)=\mathcal{C}(E)\frac{\hbar}{2\sqrt{C}E}e^{-4\pi\sqrt{C}\frac{E}{\hbar}}e^{-\pi\left(\sqrt{C}-\frac{ B}{2\sqrt{A}}\right)\frac{E}{\hbar}}~.
\end{equation}
Here we use $E=\hbar \omega$ as we consider a massless scalar field. The expression of reflection coefficient shows an exponential fall of with energy which is quite natural. Energy of the particle is much higher than the height of the potential so the probability of reflection should be extremely small \cite{griffiths:quantum}.    Now
for a large energy plank black body distribution can be approximated as proportional to the $e^{-\beta E}$ where $\beta=\frac{1}{k_BT}$. So comparing the exponential term of the expression of reflection coefficient $\mathcal{R}(\omega)$ with $e^{-\beta E}$ we can argue that the background spacetime has a temperature
\begin{equation}\label{eq22}
T=\frac{1}{\frac{4\pi}{\hbar}\sqrt{C}+\frac{\pi}{\hbar}\left(\sqrt{C}-\frac{ B}{2\sqrt{A}}\right)}
\end{equation}
We have calculate the temperature from the Boltzmann distribution part of $R(\omega)$. The energy dependent coefficient front of Boltzmann distribution could be consider as a $log$ correction to the temperature.  In section \ref{exactsolution} we perform an exact calculation of the reflection coefficient for an approximated potential the results show that the reflection coefficient follows a purely M-B distribution without any energy-dependent coefficient. 
\subsubsection{Temperature of Schwarzschild black hole}

For a Schwarzschild black hole we could calculate $A$ $B$ and $C$ explicitly from Schwarzschild metric. The line element of the Schwarzschild metric is given below \cite{Schwarzschild:1916ae},
\begin{equation}
ds^2=-\bigg(1-\frac{2M}{r}\bigg)dt^2+\frac{1}{\bigg(1-\frac{2M}{r}\bigg)}dr^2+r^2d\Omega^2~.
\end{equation}
Compering with the metric (\ref{eqn:3}) we get $f(r)=\bigg(1-\frac{2M}{r}\bigg)$ and
\begin{align*}
\sqrt{C}-\frac{B}{2\sqrt{A}}=&\frac{1}{2\kappa}-\left(-\frac{1}{8\kappa^3}f^{\prime\prime}(r_0)\right)\times\\&\frac{1}{2\sqrt{\left(\frac{1}{4\kappa^2}\left(\frac{3f^{\prime\prime}(r_0)^2}{16\kappa^2}-\frac{f^{\prime\prime\prime}(r_0)}{6\kappa}\right)\right)}}\\=&\frac{1}{2\kappa}-\frac{1}{8\kappa^3}\frac{1}{2M^2}\frac{4\sqrt{3}\kappa^2}{\sqrt{9\frac{1}{4M^4}-24\kappa\frac{1}{4 M^3}}}\\=&2M-2M=0~.
\end{align*}
Now from Eqn.(\ref{eq22}), we can calculate temperature for a Schwarzschild black hole as,
\begin{equation}
    T_{BH}=\frac{\hbar\kappa}{2\pi}~.
\end{equation}
We show that
the temperature matches with conventional Hawking temperature for Schwarzschild black
holes.

\subsubsection{Temperature of non-extremal Reissner–Nordström black hole}

\begin{figure}[h!]
\includegraphics[width=5cm]{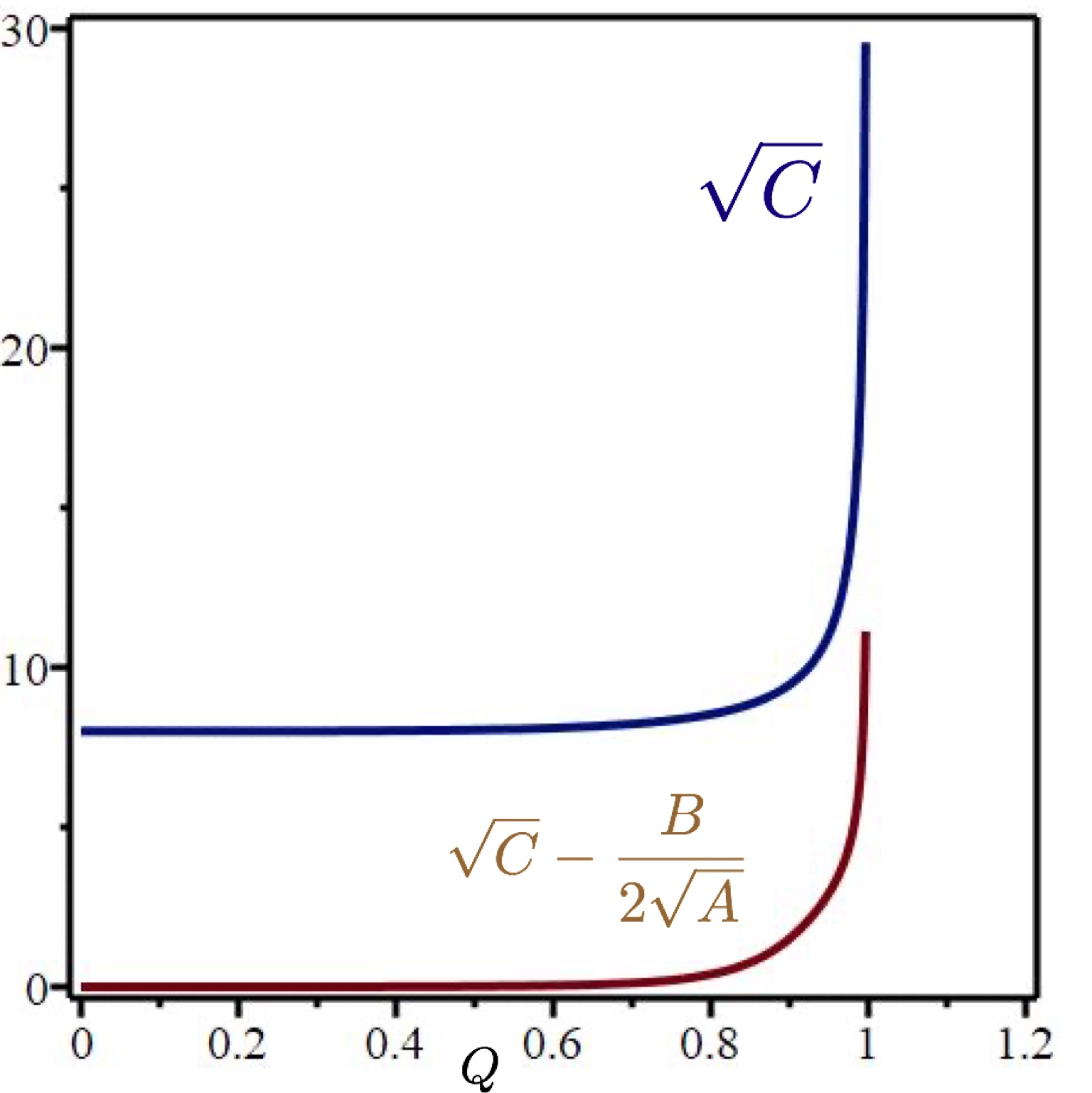}
\caption{ Here we have plotted $\sqrt{C}$ and $\sqrt{C}-\frac{B}{2\sqrt{A}}$ vs. $Q$ . Here we have taken $M=1$ where $M=Q=1$ is the extremal limit.
}
\label{figure1} 
\end{figure}

To calculate the temperature of non-extremal Reissner–Nordström black hole we follow the same procedure. Metric of a charged black hole is given as follows \cite{reissner_h_1916_1447315,book:15209}, 
\begin{equation}
ds^2=-\bigg(1-\frac{2M}{r}+\frac{Q^2}{r^2}\bigg)dt^2+\frac{1}{\big(1-\frac{2M}{r}+\frac{Q^2}{r^2}\big)}dr^2+r^2d\Omega^2~,
\end{equation}
where $M$ and $Q$ are the mass and electric charge of the black hole. Compering the above equation with Eqn. (\ref{eqn:3}) we get  
$f(r)=\left(1-\frac{2M}{r}+\frac{Q^2}{r^2}\right)$. The radius of the Horizons is computed from the equation $f(r)=0$. It  gives the  radius of the event Horizon $(r_0)$ as, $r_0=M+\sqrt{M^2-Q^2}$. As Reissner–Nordström metric is a spherically symmetric metric we could express surface gravity for the event horizon in a similar way \emph{i.e.} $\kappa=\frac{1}{2}f^{\prime}(r_0)$. So the first term in the denominator of the expression of the temperature (\ref{eq22}) gives the exact Hawking temperature but the second term in the denominator gives a nonzero value, unlike the Schwarzschild solution. An explicit straight forward calculation shows that for a non-extremal charged black hole second term $\frac{\pi}{\hbar}(\sqrt{C}-\frac{B}{2\sqrt{A}})$ is very small compare to the first term $\frac{\pi}{\hbar}\sqrt{C}$. We also show that in the FIG. \ref{figure1}. So here, we show that the temperature matches conventional Hawking temperature with a small correction. 

\subsubsection{Temperature of extremal Reissner–Nordström black hole}

Metric of an extremal Reissner–Nordström black hole is given by \cite{reissner_h_1916_1447315},
\begin{equation}
ds^2=-\bigg(1-\frac{M}{r}\bigg)^2dt^2+\frac{dr^2}{\bigg(1-\frac{M}{r}\bigg)^2}+r^2d\Omega^2~.
\end{equation}
Field equation with near horizon approximation can be written as, 
\begin{equation}\label{eq27}
\frac{d^2\phi}{d\epsilon^2}+\omega^2\frac{(\epsilon+M)^4}{\epsilon^4}\phi=0~.
\end{equation}
General solution of equation (\ref{eq27}) is a linear combination of Heun function 
\begin{equation}
\begin{split}
&\phi \left( \epsilon \right) ={\it \_C1}\sqrt {
	\epsilon}\,{{\rm e}^{{\frac {i \left( -
				\epsilon+M \right) \omega\, \left( \epsilon+M \right) }{\epsilon}}}}\times\\&{
	\it HeunD} \left( 8\,i\omega\,M,-16\,{\omega}^{2}{M}^{2}-1,0,-48\,{
	\omega}^{2}{M}^{2}+1,{\frac {\epsilon-M}{\epsilon+M}} \right)\\+&{\it \_C2}\sqrt {
	\epsilon}\,{{\rm e}^{{\frac {-i \left( -\epsilon+M \right) 
				\omega\, \left( \epsilon+M \right) }{\epsilon}}}}\times\\&{\it HeunD} \left( -8
\,i\omega\,M,-16\,{\omega}^{2}{M}^{2}-1,0,-48\,{\omega}^{2}{M}^{2}+1,{
	\frac {\epsilon-M}{\epsilon+M}} \right)~.
\end{split}
\end{equation}
To calculate the reflection coefficient, we need the asymptotic expansion of the Heun function. But the nature of the asymptotic form of Heun function is not well established in the mathematics literature. So to avoid the Heun solution, we approximate the differential equation as 
\begin{equation}\label{eq29}
\frac{d^2\phi}{d\epsilon^2}+\omega^2\bigg(\frac{6M^2}{\epsilon^2}+\frac{4M^3}{\epsilon^3}+\frac{M^4}{\epsilon^4}\bigg)\phi=0~.
\end{equation}
Solution of equation (\ref{eq29}) could be express as linear combination of Confluent Hypergeometric function.
\begin{equation}
\begin{split}
\phi=&C_1e^{-\frac{iM^2\omega}{\epsilon}}\bigg(\frac{iM^2\omega}{\epsilon}\bigg)^{i\sqrt{6}M\omega+\frac{1}{2}}\times\\&\text{U}\bigg(\frac{1}{2}+2iM\omega+i\sqrt{6}M\omega,1+i2\sqrt{6}M\omega,\frac{2iM^2\omega}{\epsilon}\bigg)\\+&C_2e^{-\frac{iM^2\omega}{\epsilon}}\bigg(\frac{iM^2\omega}{\epsilon}\bigg)^{i\sqrt{6}M\omega+\frac{1}{2}}\times\\&\text{M}\bigg(\frac{1}{2}+2iM\omega+i\sqrt{6}M\omega,1+i2\sqrt{6}M\omega,\frac{2iM^2\omega}{\epsilon}\bigg)~.
\end{split}
\end{equation}
Using asymptotic expansion of Confluent Hypergeometric function we could re express the above solution as 
\begin{equation}
\begin{split}
\phi=&C_1 e^{-\frac{iM^2\omega}{\epsilon}}\bigg(\frac{iM^2\omega}{\epsilon}\bigg)^{-2iM\omega}\\&+C_2\frac{\Gamma\bigg(1+i2\sqrt{6}M\omega\bigg)}{\Gamma\bigg(\frac{1}{2}+2iM\omega+i\sqrt{6}M\omega\bigg)}e^{\frac{iM^2\omega}{\epsilon}}\bigg(\frac{iM^2\omega}{\epsilon}\bigg)^{2iM\omega}~.\end{split}
\end{equation}
Now using the same procedure as discussed above, we could calculate the reflection coefficient as follows
\begin{eqnarray}
R(\omega)=\frac{\Gamma\bigg(1+i2\sqrt{6}M\omega\bigg)}{\Gamma\bigg(\frac{1}{2}+2iM\omega+i\sqrt{6}M\omega\bigg)}\times\nonumber\\ \frac{\Gamma\bigg(1-i2\sqrt{6}M\omega\bigg)e^{-4\pi M\omega}}{\Gamma\bigg(\frac{1}{2}-2iM\omega-i\sqrt{6}M\omega\bigg)}~.
\end{eqnarray}
After applying the identities (\ref{identities}), one can simplify the reflection coefficient as,
\begin{equation}
\begin{split}
R(\omega)&=\frac{\pi 2\sqrt{6}M\omega}{\sinh\pi(2\sqrt{6}M\omega)}\frac{\cosh \pi\bigg(2M\omega+\sqrt{6}M\omega\bigg)}{\pi}e^{-4\pi M\omega}\\
&=2\sqrt{6}M\omega e^{-2\pi M\omega-\sqrt{6}\pi M\omega}\\
&=2\sqrt{6}M\omega e^{-4.45\pi M\omega}\\
&=\frac{2\sqrt{6}M E}{\hbar} e^{-4.45\pi M\frac{E}{\hbar}}~.
\end{split}
\end{equation}
Here we use $E=\hbar \omega$ as we consider a massless scalar field.
Now from the above equation, we can calculate the temperature for an extremal Reissner–Nordström black hole as,
\begin{equation}
T_{BH}=\frac{\hbar}{4.45\pi M}~.
\end{equation}
Here we have got a non-vanishing  Hawking temperature for an extremal Reissner–Nordström black hole.

\section{Exact solution using an approximated potential} \label{exactsolution}
\subsection{For Schwarzschild spacetime}
In privious section we get Schrodinger like equation (\ref{eqn:5}) in $r_*$ coordinate. But we can't solve it in $r_*$ coordinate as we cant express $r$ in terms of $r_*$ so the potential $V(r(r_*)))=\frac{f'(r(r_*))f(r(r_*))}{r}$. To avoid this problem we try to solve the equation (\ref{eqn:5}) in $r$ coordinate but in $r$ coordinate the solution is a linear combination of Heun function like for Schwarzschild spacetime  it would looks like 
\begin{equation}
	\begin{split}
	&\phi \left( r \right) ={\it C1}\,{{\rm e}^{i\omega\,r}} \left( -r+2\,m
	\right) ^{2\,i\omega\,m}\times\\
	&{\it HeunC}
	\left( -4\,im\omega,4\,i\omega\,m,0,-8\,{\omega}^{2}{m}^{2},8\,{
		\omega}^{2}{m}^{2},1/2\,{\frac {-r+2\,m}{m}} \right)\\&+{\it C2}\,{{\rm e}^{\omega\,ri}}\left( -r+2\,m \right) ^{-2\,i\omega\,m}\times\\
	&{\it HeunC} \left( -4\,im\omega,-4\,im
	\omega,0,-8\,{\omega}^{2}{m}^{2},8\,{\omega}^{2}{m}^{2},1/2\,{\frac {-
			r+2\,m}{m}} \right)~.
	\end{split}
\end{equation} 
As we discussed previously, the Heun function's asymptotic nature is not well known to us till now, we go for near horizon approximation. But an objection may arise that this quantum mechanical reflection occurs due to the near horizon approximation of potential. So in this section, we do an exact calculation without doing near horizon approximation but with approximated potential that nearly matches with $V(r(r_*))$ for the entire space-time. Although we can't express the potential $V(r(r_*))$ explicitly in $r_*$ coordinate but we can make a plot of the potential in $r_*$ coordinate using $Lambart W$ function, and from there, we could easily figure out it nearly matches with  The Poschl-Teller potential which is given by
\begin{equation}\label{eq36}
	V_{PT}=\frac{U_0\alpha^2}{\cosh^2{(\alpha r_*)}}~.
\end{equation}
For Schwarzschild spacetime expression of potential in $r$ coordinate is 
\begin{equation}\label{eq37}
	V(r(r_*))=\frac{2M}{r(r_*)^3}\bigg(1-\frac{2M}{r(r_*)}\bigg)~.
\end{equation}
To write the $r$ in terms of $r_*$ we solve the the equation $r_*=r+ 2M\textrm{ln}(r-2M)$ using $LambertW$ function ( for $LambertW$ function,  see Appendix \ref{LambertW}) as,
\begin{equation}
	r(r_*)=2M\bigg(LambertW\bigg(\frac{e^{\frac{r_*-2M}{2M}}}{2M}\bigg)+1\bigg)~. 
\end{equation}  
Now using the expression of $r(r_*)$ in equation (\ref{eq37}) we could make a plot of potential in $r_*$ coordinate for any rage of $r_*$ between $r_*\in (-\infty,\infty)$ which nearly matches with the expression (\ref{eq36}) with a shift along $r_*$ and that can be fix by redefine $r_*$ as $r_*=r+2M\textrm{ln}(r-2M)+2M\textrm{ln}\frac{3}{2}-\frac{8M}{3}$. We could fix the value of alpha by demanding first derivative of $V_{PT}$ at horizon is nonzero also we could calculate the value of $U_0$ by demanding maximum value of $V_{PT}$ should be same with maximum of $V(r)$ (see Appendix \ref{Calculation of alpha}). Applying all these condition we get the value of alpha and $U_0$ as
\begin{equation}
	\alpha=\frac{1}{4M}\;\;\;\textrm{and}\;\; U_0 =\frac{27\times4M}{1024 M^2}~.
\end{equation}   
\begin{figure}[h!]
	\includegraphics[width=5cm]{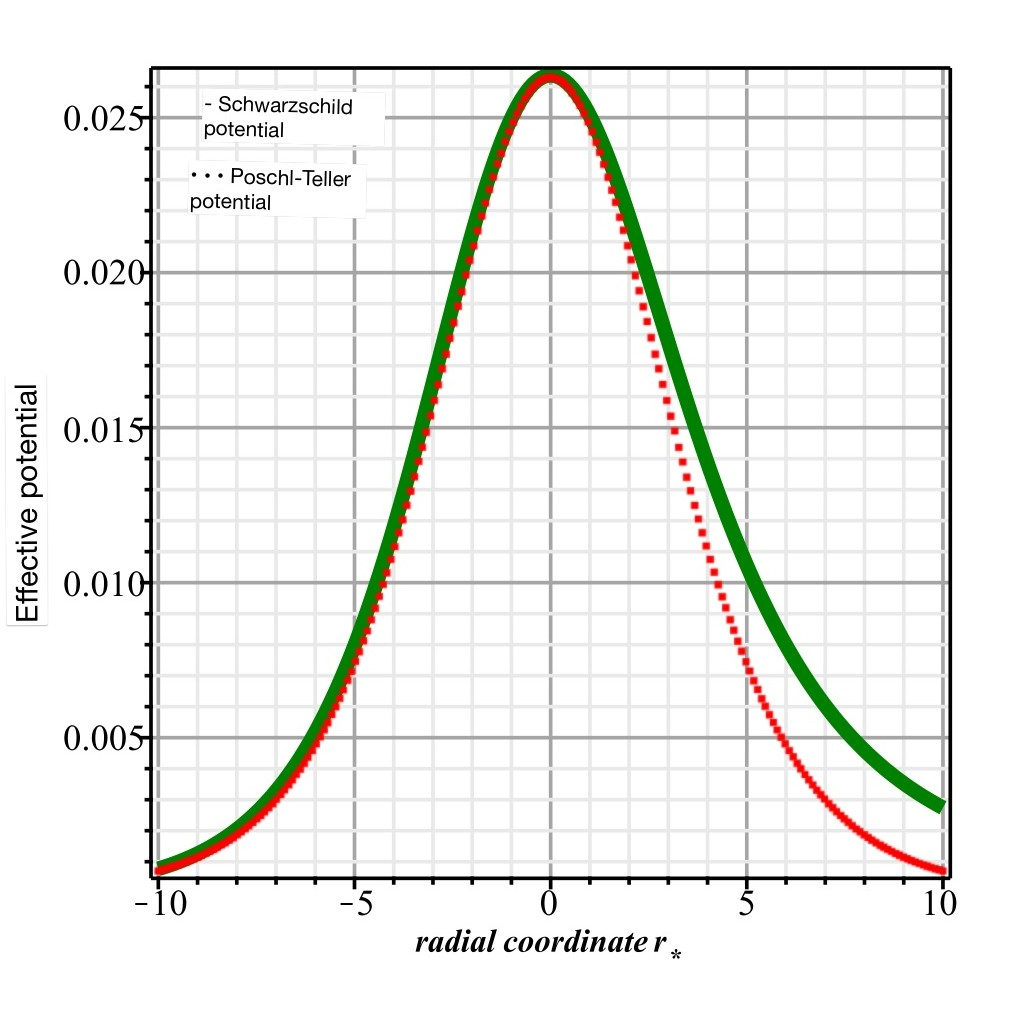}
	\caption{ Here we have plotted $V_{PT}$ and $V(r_*)$ vs. $r_*$ . Here we have taken $M=1$.
	}
	\label{figure2} 
\end{figure}
 We use this potential in one dimensional Schrodinger equation (\ref{eqn:5}) to derive the reflection coefficient. Using the derivation of the book (\cite[p.~80]{book:14848}) one can get the reflection coefficient as,
\begin{align*}
	R=&\frac{1}{1+\sec^2{[\frac{\pi}{2}\sqrt{(1-8m U_0/\hbar^2)}]}\sinh^2{(\frac{\pi\kappa}{\alpha})}}\\\simeq& \cos^2{[\frac{\pi}{2}\sqrt{(1-8m U_0/\hbar^2)}]}\left(\frac{2}{e^{(\frac{\pi\kappa}{\alpha})}-e^{(-\frac{\pi\kappa}{\alpha})}}\right)^2\\\simeq& 4\cos^2{[\frac{\pi}{2}\sqrt{(1-8m U_0/\hbar^2)}]}e^{-\frac{2\pi\kappa}{\alpha}}~.
\end{align*}
Now let us consider $A=4\cos^2{[\frac{\pi}{2}\sqrt{(1-8m U_0/\hbar^2)}]}$ and for using $\kappa=\frac{\sqrt{2}\omega}{\hbar}$ the reflection coefficient can be written as,
\begin{equation}
	R=A e^{-\frac{2\sqrt{2}\pi}{\hbar\alpha}\omega}~.
\end{equation}
By comparison with Planck distribution and using the value of $\alpha=\frac{1}{4M}$, the corresponding temperature is then given by,
\begin{equation}
	T=\frac{\hbar}{8\sqrt{2}\pi M}~.   
\end{equation}

\subsection{For Reissner-Nordström spacetime}

Here we want to calculate the exact solution of the reflection coefficient using the approximated potential for Reissner- Nordström spacetime. But due to the presence of two horizons, two logarithmic terms arrive when we express $r_{\star}$ in terms of $r$ which is given by, 
\begin{equation}
	r_*=r+\frac{1}{2\kappa_+}\textrm{log}(r-r_+)+\frac{1}{2\kappa_-}\textrm{log}(r-r_-)~.
\end{equation}
Due to the presence of different coefficients in the front of the logarithmic term, we get generalized LambertW function when we want to express $r$ in terms of $r_{\star}$. This generalized LambertW function is not easy to handle. Thus we can't express $V(r(r_{\star}))$ for the entire spacetime. But if we study the dependence of $r_*$ on $r$ near-horizon region and for a large distance from the outer horizon then the relation $(42)$, looks like, 
\begin{equation}
	\begin{split}
	r_*&\approx \frac{1}{2\kappa_+}\textrm{ln}(r-r_+)\;\;\;\;\;\;\;\;\;\;r\rightarrow r_+ \\
	&\approx r+\bigg(\frac{1}{2\kappa_+}+\frac{1}{2\kappa_-}\bigg)\textrm{ln}(r) \;\;\; r\gg r_+~.
	\end{split}
\end{equation}
Then there are only one logarithmic term presences in $r_{\star}$ coordinate. Now we could express $r$ in terms of  $r_*$ using the $LambertW$ function ( for $LambertW$ function,  see Appendix \ref{LambertW}) for both regions. We had seen in the previous section that the plot of potential in $r_*$ coordinate matches with the profile of $\textrm{Sech}^2(\alpha r_*)$. The nature of the potential for Reissner-Nordström spacetime in $r$ coordinate is similar to the potential of Schwarzschild spacetime, and the expression of $r$ in terms of $r_*$ is similar at least for the two regions of spacetime so that we could expect the potential for Reissner-Nordström spacetime matches with the profile of  $\textrm{Sech}^2(\alpha r_*)$. Now using a similar procedure, we could calculate the value of $\alpha$ as (see Appendix \ref{Calculation of alpha}),
\begin{equation}
	\alpha=\kappa_+~.
\end{equation}   
Then we can calculate the reflection coefficient similarly as we discussed in the previous section. The reflection coefficient becomes
\begin{equation}
	R=A e^{-\frac{2\sqrt{2}\pi}{\hbar\kappa_+}\omega}~.
\end{equation}
So the temperature is given by,
\begin{equation}
	T=\frac{\hbar\kappa_+}{2\sqrt{2}\pi }~.   
\end{equation}

\section{Conclusion}\label{conclusion}

In this paper, we have explored an alternative derivation of the Hawking temperature and radiation. We have used the analogy with a simple quantum mechanical reflection phenomenon from a one-dimensional potential. The reflection probability exhibits an exponential fall-off in energy which indicates a close analogy with a finite temperature Boltzmann probability distribution at a finite temperature. This strategy has been applied to the cases of Schwarzschild and Reissner-Nordström black holes. We have shown that the temperature, obtained in this method, matches exactly with the Hawking temperature for a Schwarzschild black hole. For a non-extremal Reissner-Nordström black hole, this temperature closely matches with the Hawking temperature up to a multiplicative factor, which is found to sharply approach unity for near-extremal RN black holes. We have also found that a non-vanishing Hawking temperature arises for an extremal RN black hole in this approach \cite{Hawking:1994ii}. This is an exact local calculation of the Hawking temperature. However, it has one crucial difference from the tunneling approach that in this case, it is not necessary to involve regions inside the black hole horizon.

To arrive at a single particle quantum mechanical model, our starting point has been the field equation. As a result, in the analog QM equation, the energy $E$ appears as square, $E^2$. It produces the expected exponential fall-off with energy $E$ when we use the quantum mechanical reflection calculations. This shows that a naive quantum mechanical model will not give the same result and the correct fall-off with energy. We have to rely on field equations to arrive at the correct quantum mechanical model. The deviations in temperatures of non-extremal and extremal Reissner-Nordström black holes from other field theoretic calculations are characteristic features of the present calculations. Many authors have pointed out that temperatures of extremal black holes may not be correctly represented by their surface gravity. A nonzero temperature from this approach provides an interesting guideline for the deviation of temperatures from classical surface gravity for extremal black holes.       

In calculating the reflection coefficient, we have used two approximation methods. In the first method, we have approximated the exact potential by making a near-horizon expansion truncated at fourth order (up to which a reliable asymptotic behavior can be extracted) and then calculated the reflection coefficient. In the second method, we have taken a simulated a potential that fits the exact potential over a wide region of space and  then calculated the reflection coefficient exactly from the simulated potential. The apparently two distinct approximations give temperatures that differ by a $o(1)$ factor of $\sqrt{2}$. This fact that two apparently distinct approximations give similar temperatures, give strong support in favour of the approximations used in calculations.

In this paper, we have considered only spherically symmetric black holes. It would be nice to apply this formalism axisymmetric black hole like Kerr black hole. 

\begin{acknowledgments}
CS thanks the Saha Institute of Nuclear Physics (SINP) Kolkata for financial support.
\end{acknowledgments}

\section*{Data Availability}
My manuscript has no associated data

\appendix

\section{Confluent Hypergeometric Functions}\label{appendix}

Let us consider a second order differential equation of the form
\begin{equation}\label{eqA1}
z\frac{d^2w}{dz^2}+(b-z)\frac{dw}{dz}-aw=0~.
\end{equation}
This differential equation is known as confluent Hypergeometric differential equation or sometimes called Kummer's differential equation \cite{book:10383,NIST:DLMF}. This equation has a regular singular point at 0 and irregular singular point at $\infty$.\par
By substituting $W=e^{-\frac{1}{2}z}z^{\frac{1}{2}+\mu}w,\kappa=\frac{1}{2}b-a$, and $\mu=\frac{1}{2}b-\frac{1}{2}$
into equation (\ref{eqA1}) we will obtain the differential equation as follows,
\begin{equation}
\frac{d^2W}{dz^2}+\left(-\frac{1}{4}+\frac{\kappa}{z}+\frac{\frac{1}{4}-\mu^2}{z^2}\right)W=0~.
\end{equation}
This equation is known as Whittaker's differential equation. The solutions of this equation is known as Whittaker functions , $W_{\kappa\mu}(z)$. In terms of Confluent Hypergeometric functions M and U, the Whittaker functions are:
\[M_{\kappa\mu}(z)=e^{-\frac{1}{2}z}z^{\frac{1}{2}+\mu}M\left(\frac{1}{2}+\mu-\kappa,1+2\mu,z\right)~,\]
\[U_{\kappa\mu}(z)=e^{-\frac{1}{2}z}z^{\frac{1}{2}+\mu}W\left(\frac{1}{2}+\mu-\kappa,1+2\mu,z\right)~.\]
In our calculation equation (\ref{eq11}) is the Whittaker's differential equation whose solution is equation (\ref{eq13}).
The limiting forms of Kummer's functions are given by 
for $ z\rightarrow0$\par
\[M(a,b,z)=1+\mathcal{O}(z)~,\]
\[U(a,b,z)=\frac{\Gamma(b-1)}{\Gamma(a)}z^{1-b}~.\]

\section{LambertW function}\label{LambertW}

\begin{figure}[h!]
	\includegraphics[scale=.4]{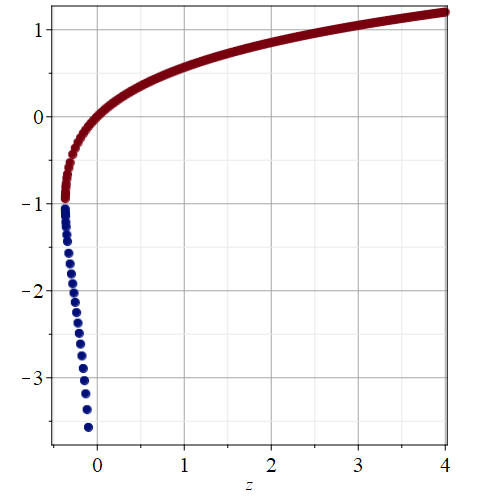}
	\caption{ Here We have plotted two real branches of LambertW function,$W_0(z)$ and $W_{-1}(z)$ vs. z.
	}
	\label{figure3} 
\end{figure}

In mathematics and physics we often encounter with different transcendental equations. 
The LambertW function, also called the omega function is the solution of the equation \cite{article}
\begin{equation}\label{eq4}
	we^{w}=z~,
\end{equation}
where z is any complex number. The equation has countably infinite number of solutions, denoted by $W_k(z)$. Where k is an integer and specifies a branch of Lambert W function. There are only two real-valued branches, denoted as $W_0(z)$ and $W_{-1}(z)$. $W_0(z)$ is called the principal branch. $W_0(z)\geq-1$ and $W_{-1}(z)\leq-1$ ,where $-\frac{1}{e}\leq z <0$ (FIG. \ref{figure3}). When $z=-\frac{1}{e}$ then the equation \ref{eq4} has a double root $w=-1$ and this implies,
\[W_0(-{1}/{e})=W_{-1}(-{1}/{e})=-1~.\]

\begin{widetext}
\section{Calculation of $\alpha$}\label{Calculation of alpha}

First derivative of $V(r)$ at the horizon $r=2M$ is nonzero so the first derivative of $V_{PT}$ at $r=2M$ should be non-zero in $r$ coordinate for Schwarzschild spacetime. Now
\begin{equation}
	\frac{V_{PT}}{dr}=\frac{V_{PT}}{dr_*}\frac{dr_*}{dr}~,
\end{equation}

for Schwarzschild spacetime can be expressed as,
\begin{equation}
	\begin{split}
	\frac{V_{PT}}{dr}&=(r-2M)^{4M\alpha-1}\times \\
	&\frac{\textrm{sinh}(\alpha r)\left\lbrace \right(r-2M)^{4M\alpha}+1\rbrace +\textrm{cosh}(\alpha r)\left\lbrace \right(r-2M)^{4M\alpha}-1\rbrace  }{\textrm{cosh}(\alpha r)\left\lbrace \right(r-2M)^{4M\alpha}+1\rbrace +\textrm{sinh}(\alpha r)\left\lbrace \right(r-2M)^{4M\alpha}-1\rbrace}~.
	\end{split}
\end{equation}
So to get a non zero divergence free first derivative at $r=2M$
\begin{equation}
	4M\alpha-1=0\;\;\ \textrm{or}\;\;\;\ \alpha=\frac{1}{4M}~.
\end{equation}
Similarly for Reissner-Nordström spacetime, 
\begin{equation}
	\begin{split}
	&\frac{V_{PT}}{dr}=(r-r_+)^{\frac{\alpha}{\kappa_+}-1}(r-r_-)^{\frac{\alpha}{\kappa_-}-1}\times \\
	&\frac{\textrm{sinh}(\alpha r)\left\lbrace \right(r-r_+)^{\frac{\alpha}{\kappa_+}}(r-r_-)^{\frac{\alpha}{\kappa_-}}+1\rbrace +\textrm{cosh}(\alpha r)\left\lbrace \right(r-r_+)^{\frac{\alpha}{\kappa_+}}(r-r_-)^{\frac{\alpha}{\kappa_-}}-1\rbrace  }{\textrm{cosh}(\alpha r)\left\lbrace \right(r-r_+)^{\frac{\alpha}{\kappa_+}}(r-r_-)^{\frac{\alpha}{\kappa_-}}+1\rbrace +\textrm{sinh}(\alpha r)\left\lbrace \right(r-r_+)^{\frac{\alpha}{\kappa_+}}(r-r_-)^{\frac{\alpha}{\kappa_-}}-1\rbrace }~.
\end{split}
\end{equation}
For a non-zero first derivative at $r=r_+$
\begin{equation}
	\frac{\alpha}{\kappa_+}-1=0\;\;\ \textrm{or}\;\;\;\ \alpha=\kappa_+~.
\end{equation}
\end{widetext}


\nocite{*}
\bibliographystyle{ieeetr}
\bibliography{apssamp}

\end{document}